\documentclass[]{spie}  

\usepackage{amsmath,amsfonts,amssymb,caption}
\usepackage{graphicx}
\usepackage{subfig}
\usepackage[colorlinks=true, allcolors=blue]{hyperref}

\usepackage[labelsep=period]{caption}
\title{All-dielectric metasurface for wavefront control at terahertz
frequencies}

\author[a,b,c]{Raghu Dharmavarapu}
\author[a,c]{Soon Hock Ng}
\author[b]{Shanti Bhattacharya}
\author[a,c]{Saulius Juodkazis}
\affil[a]{Centre for Micro-Photonics, Faculty of Science, Engineering and Technology, Swinburne University of Technology, Hawthorn VIC 3122, Australia}
\affil[b]{Centre for NEMS and Nanophotonics (CNNP), Department of Electrical Engineering, Indian Institute of Technology Madras, Chennai 600036, India}
\affil[c]{Melbourne Centre for Nanofabrication, the Victorian Node of the Australian National Fabrication Facility, 151 Wellington Rd., Clayton 3168 VIC, Australia}

\authorinfo{Raghu Dharmavarapu. E-mail: raghu.d@ee.iitm.ac.in}

\begin{document} 
\maketitle

\begin{abstract}
Recently, metasurfaces have gained popularity due to their ability to offer a spatially varying phase response, low intrinsic losses and high transmittance. Here, we demonstrate numerically and experimentally a silicon metasurface at THz frequencies that converts a Gaussian beam into a Vortex beam independent of the polarization of the incident beam. The metasurface consists of an array of sub-wavelength silicon cross resonators made of a high refractive index material on substrates such as sapphire and CaF$_2$ that are transparent at IR-THz spectral range. With these substrates, it is possible to create phase elements for a specific spectral range including at the molecular finger printing around 10$~\mu$m as well as at longer THz wavelengths where secondary molecular structures can be revealed. This device offers high transmittance and a phase coverage of 0 to $2\pi$. The transmittance phase is tuned by varying the dimensions of the meta-atoms. To demonstrate wavefront engineering, we used a discretized spiraling phase profile to convert the incident Gaussian beam to vortex beam. To realize this, we divided the metasurface surface into eight angular sectors and chose eight different dimensions for the crosses providing successive phase shifts spaced by $\pi/4$ radians for each of these sectors. Photolithography and reactive ion etching (RIE) were used to fabricate these silicon crosses as the dimensions of these cylinders range up to few hundreds of micrometers. Large 1-cm-diameter optical elements were successfully fabricated and characterised by optical profilometry. 
\end{abstract}


\section{INTRODUCTION}
\label{sec:intro}  

Metasurfaces~\cite{yu2014flat,genevet2017recent,sautter2015active,chen2016review,yu2015high,arbabi2015dielectric} are two dimensional array of sub-wavelength spaced meta-atoms, such as V-shaped antennas~\cite{kildishev2013planar}, cylindrical disks~\cite{chong2015polarization} and nano-fin structures~\cite{khorasaninejad2016metalenses}. With careful design these resonators can support electric and magnetic dipole resonances simultaneously and render a $2\pi$ phase coverage for the transmitting light. This provides the ability to control the phase and amplitude of the light locally thus enabling wavefront engineering. Meta surfaces made of high dielectric permittivity materials such as silicon can offer low losses and high transmission efficiencies. These structures have been experimentally demonstrated for a variety of applications such as beam steering~\cite{cheng2014wave}, holography~\cite{genevet2015holographic}, lenses~\cite{khorasaninejad2016metalenses}, vortex beam generators~\cite{yang2014dielectric} and polarization beam splitters~\cite{khorasaninejad2015efficient} at infrared and optical wavelengths.

Dielectric metasurfaces offer an ideal platform for  the development of high-efficiency optical components at THz frequencies. Various planar THz componets such as lenses~\cite{chang2017demonstration,wang2015broadband}, beam splitters~\cite{niu2013experimental} and wave plates~\cite{cong2014highly,wang2015ultrathin} have been reported. So far, dielectric metasurfaces were mainly demonstrated in the visible and infrared spectral regimes. Several THz dielectric metasurfaces based lenses were reported in the literature but they operate in reflection mode~\cite{ma2016terahertz}. In this work, we developed a transmission mode vortex  beam generator using cross shaped resonators at 0.73~THz. Finite difference time domain (FDTD) simulation, fabrication, and optical characterisation are presented.

\section{Meta-atom design and dimensions}

In this section, we discuss the design parameters and selection process for the proper dimensions of silicon cross resonators that help to obtain the desired transmission and phase response.

\begin{figure}[tb]
\centering
\subfloat[]{
\label{fig:1a}
\vspace{-2cm}
\includegraphics[height = 3.5cm]{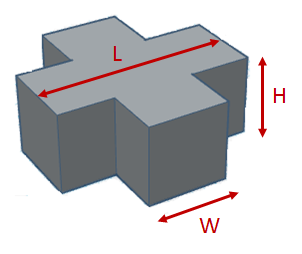}
}
\subfloat[]{
\label{fig:1b}
\includegraphics[height = 7cm]{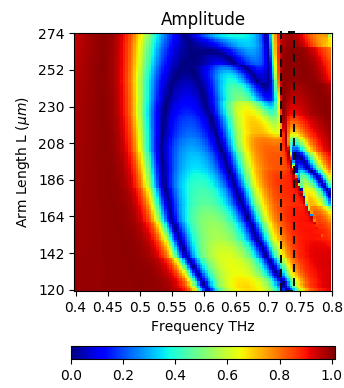}
}
\subfloat[]{
\label{fig:1c}
\includegraphics[height = 7cm]{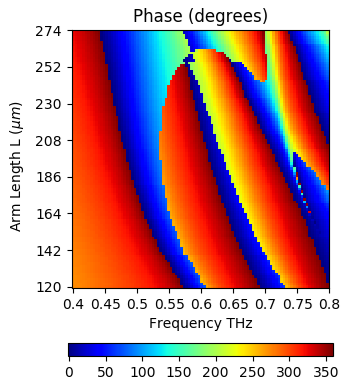}
}
\caption{
\label{fig:1}
(a) Schematic diagram of the silicon cross meta-atom with width, $W$, and height, $H$, fixed at $70~\mu$m and $150~\mu$m, respectively. (b) Simulated transmission amplitudes and (c) corresponding transmission phases as a function or arm length, $L$, over the frequency range 0.4-0.8 THz. Spwctral range enclosed between two dashed vertical lines was chosen due to constant amplitude and possibility of $2\pi$ phase control. }
\end{figure}

The schematic of the silicon cross resonator is shown in Fig.~\ref{fig:1a}. Silicon is chosen due to its high relative permittivity of $\epsilon_r = 13.5$ and its low loss nature in the terahertz regime. The width, height and lattice constant of the cross are fixed at 70$~\mu$m, 150~$\mu$m and 500~$\mu$m respectively. These dimensions are optimized to excite electric and magnetic resonances at the desired range of frequencies. We used the FDTD method to perform simulations, with a linearly polarized plane wave incident normally on the metasurface with its electric field in $y$ direction and propagating in $z$ direction. 

Due to the subwavelength spacing of the resonators and low coupling between adjacent resonators we considered the electromagnetic response of one resonator by replicating it as an infinite array in $x$ and $y$ directions achieved by the periodic boundary condition. Each resonator can support electric and magnetic resonance when the dimensions become comparable to the wavelength of the incident light~\cite{evlyukhin2010optical}. High transmission can be achieved by tuning the dimensions such that the two resonance overlap spectrally and cancel out the back scattered light. Figure~\ref{fig:1b} and \ref{fig:1c} show the transmission amplitude and phase respectively as a function of the arm length $L$ varied from 120-275 $\mu$m and frequency 0.4-0.8~THz. From the simulation data we have selected eight resonators with nearly equal phase steps of $\pi/4$ to cover full $2\pi$ range.

\begin{figure}[tb]
\centering
\subfloat[]{
\label{fig:2a}
\includegraphics[width = 7cm]{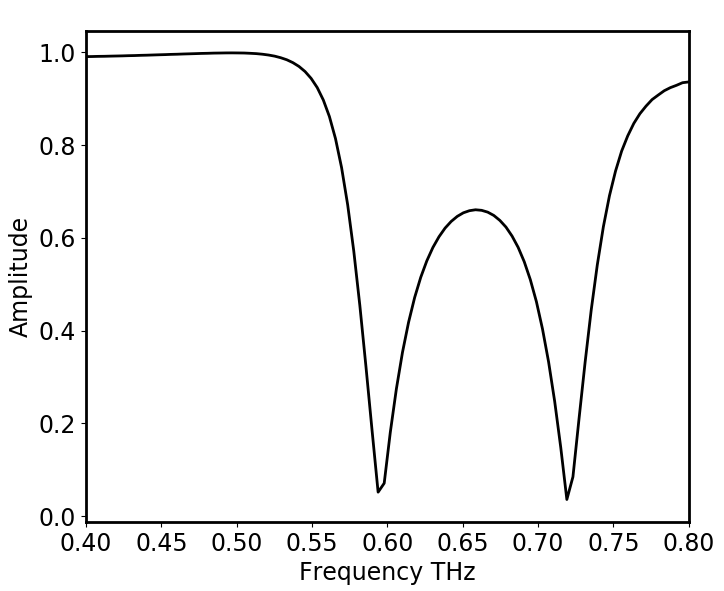}
}
\hspace{1cm}
\subfloat[]{
\label{fig:2b}
\includegraphics[width = 7cm]{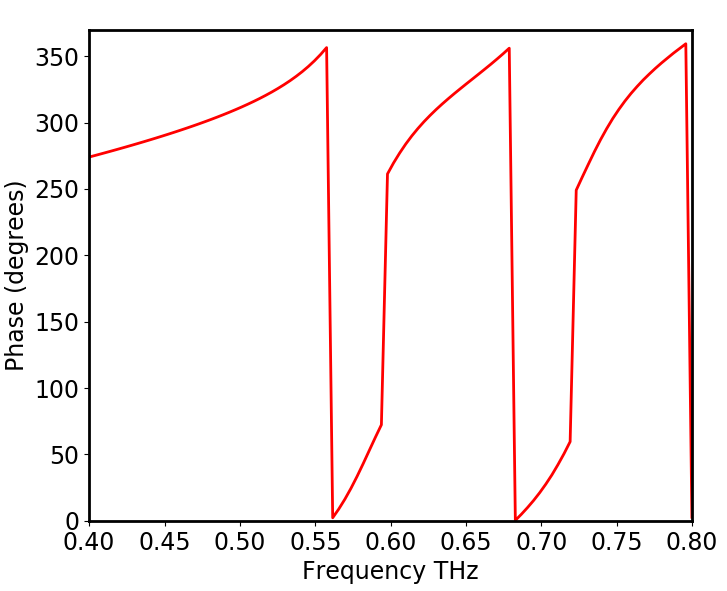}
}
\caption{
\label{fig:2}
(a) Amplitude and (b) phase of the cross resonator with arm length, $L$, of 125~$\mu$m.} 
\end{figure}

The amplitude and phase of the transmitted light is computed by using design parameters. Figure~\ref{fig:2a} and \ref{fig:2b} shows the typical amplitude and phase response across the desired frequency range 0.4-0.8~THz for a cross arm length of $L = 125~\mu$m.

\section{Vortex beam generator}

\begin{figure}[tb]
\centering
\includegraphics[width = 9cm]{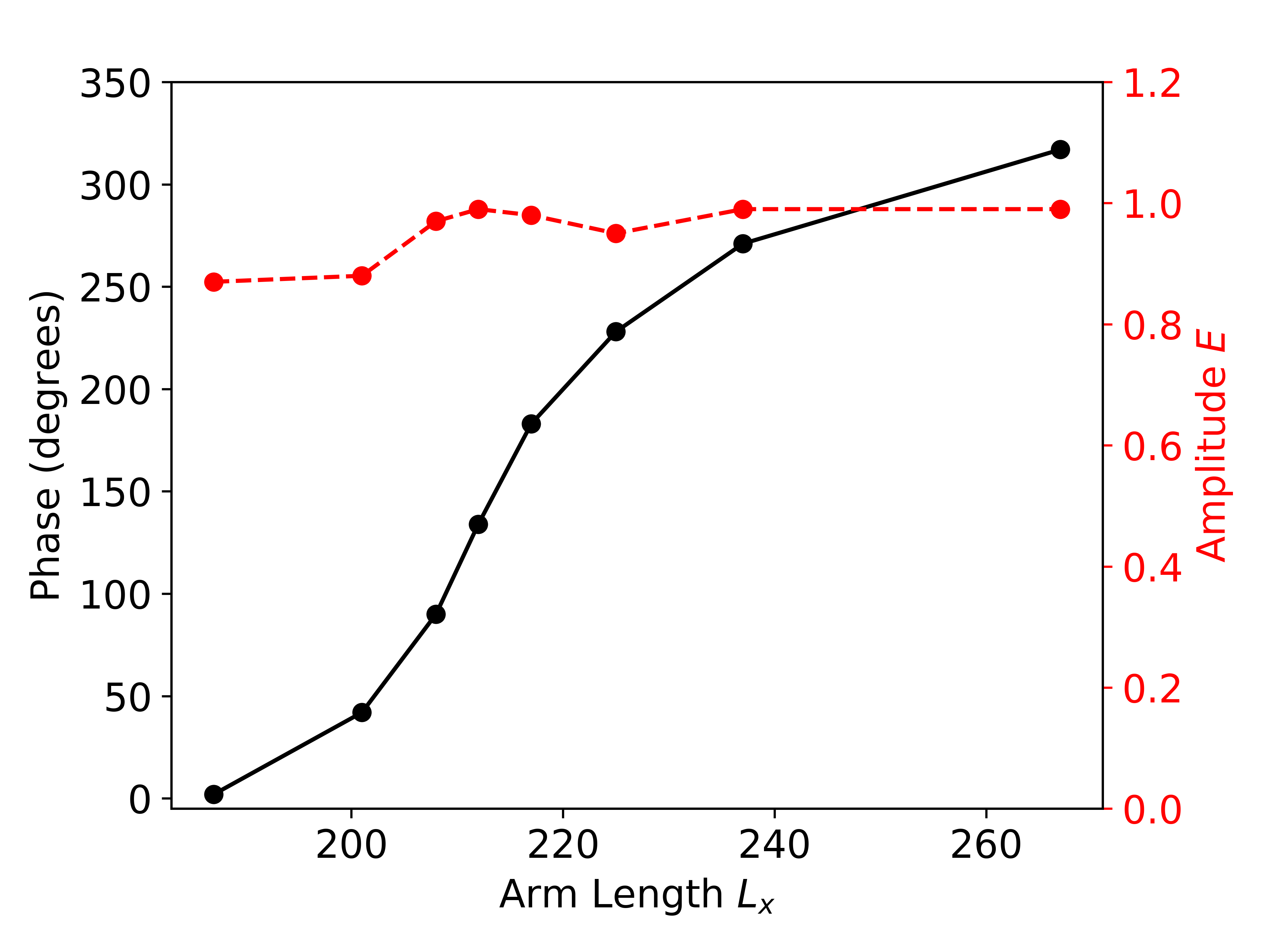}
\caption{Transmission amplitudes and phases of the cross resonators as a function of arm lengths at $f = 0.73$ THz.}
\label{fig:3}
\end{figure}

As an application, we propose a metasurface vortex beam generator at terahertz frequencies. The phase function of the vortex generator is given by a spiraling function as the following equation:
\begin{equation}
\Phi(x,y) = \tan^{-1}(y/x),
\label{eq:1}
\end{equation}
where ($x,y$) are the coordinates in the metasurface plane. The continuous phase profile of the vortex generator given in the Eq.~\ref{eq:1} is discretized to eight phase levels and the corresponding eight cross resonators dimensions are given in Table~\ref{tab:1}. The transmission phase curve is shown in Fig.~\ref{fig:3}. It is revealed that all the resonators are showing high transmission with a mean variation of 5\%.
\begin{table}
\centering
\caption{Transmission amplitudes and phases of the eight cross resonators.}
\label{tab:1}
\begin{tabular}{|l|l|l|l|l|l|l|l|l|}
\hline
\textbf{Arm Length} ($\mu$m) & 187  & 201  & 208  & 212  & 217  & 225  & 237  & 267  \\ \hline
\textbf{Amplitude  }          & 0.87 & 0.88 & 0.97 & 0.99 & 0.98 & 0.95 & 0.99 & 0.99 \\ \hline
\textbf{Phase }               & 2    & 42   & 90   & 135  & 183  & 228  & 271  & 317  \\ \hline
\end{tabular}
\end{table}
\begin{figure}[tb]
\centering
\subfloat[]{
\label{fig:4a}
\includegraphics[height = 5cm]{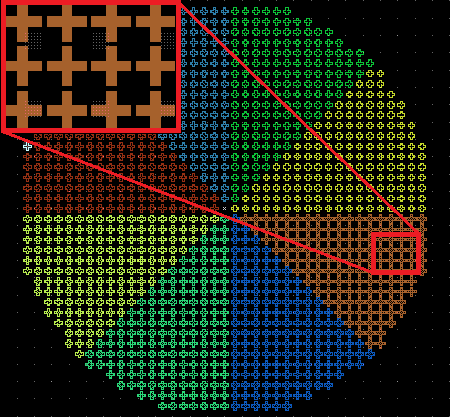}
}
\hspace{1cm}
\subfloat[]{
\label{fig:4b}
\includegraphics[height = 5cm]{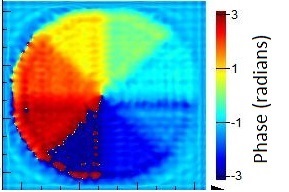}
}
\caption{
\label{fig:4}
(a) Schematic layout of meta spiral phase plate (b) Transmission phase of the electric field simulated using FDTD.} 
\end{figure}

We designed a spatial phase element with a diameter 1.2~cm consisting of eight types of cross resonators in corresponding eight octants as shown in Fig.~\ref{fig:4a}. Such optical element has dimensions useful for range of applications in THz experiments including synchrotron beamlines. The FDTD simulated transmission the phase is shown in Fig.~\ref{fig:4b}. The designed spiraling phase was indeed following the required $2\pi$ azimuthal span. 

\subsection{Fabrication}

\begin{figure}[tb]
\centering
\subfloat[]{
\label{fig:5a}
\includegraphics[height = 5cm]{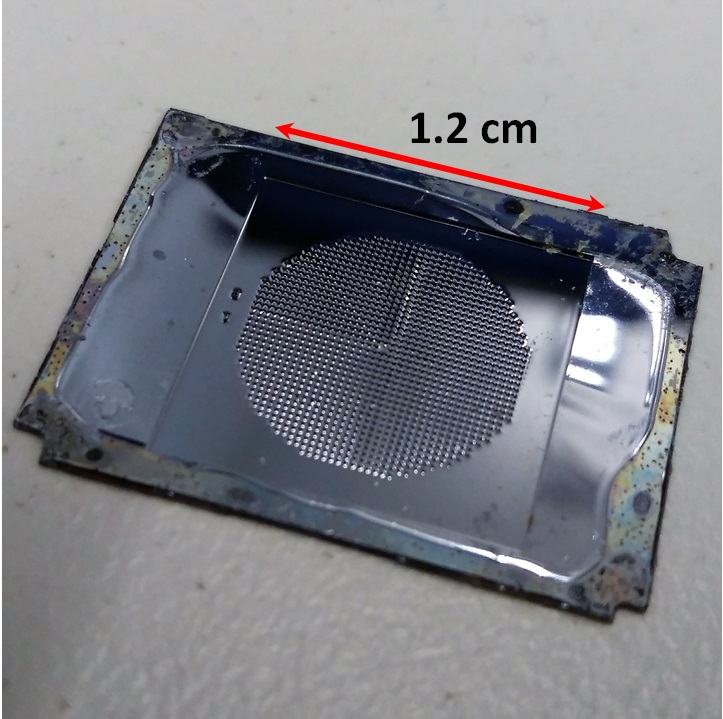}
}
\hspace{1cm}
\subfloat[]{
\label{fig:5b}
\includegraphics[height = 5cm]{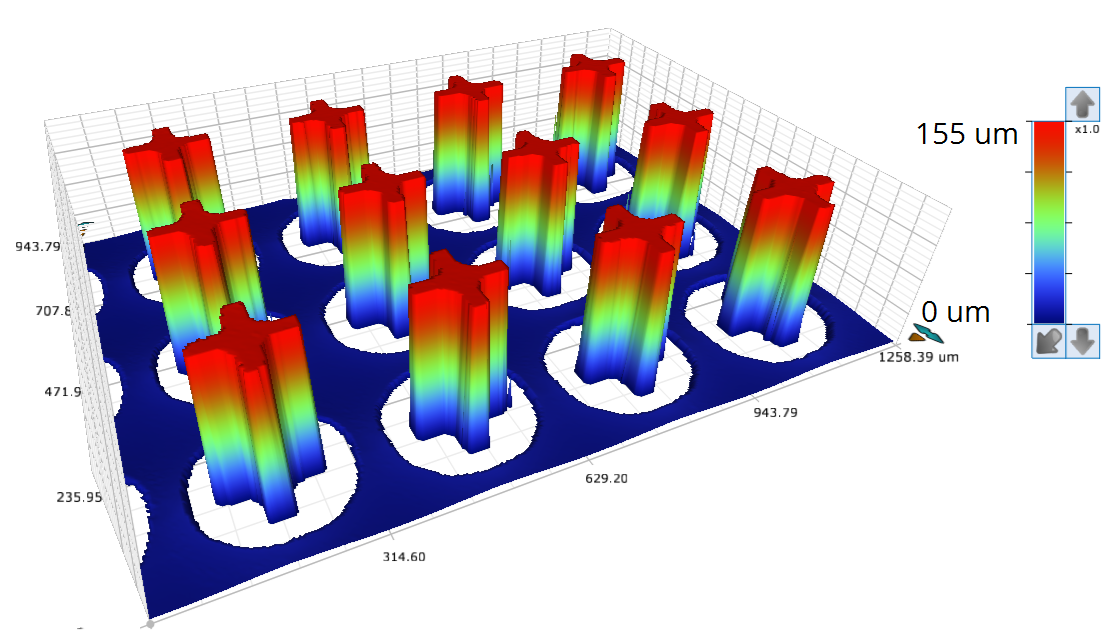}
}
\caption{
\label{fig:5}
(a) Fabricated metadevice  (b) Profile measured using an optical profilometer.} 
\end{figure}

A silicon wafer of standard 500~$\mu$m thickness was used for the fabrication of the optical meta-surface element. The device was fabricated using photolithography followed by reactive ion etching (RIE). The mask for the photolithography was fabricated using the Intelligent micropatterning SF100 XPRESS direct writing system. We used AZ4562 photo-resist in order to obtain thicker resist coating. The large thickness  was needed to obtain a high etch depth of 150~$\mu$m by RIE (the resist was a acting as a sacrificial mask). The standard Bosh process was used to etch silicon for the required high aspect ratio pattern. A photograph of the fabricated metadevice and the optical profilometer measurement are shown in Fig.~\ref{fig:5a} and \ref{fig:5b}. 
The required depth of 150~$\mu$m was achieved after 90~min plasma etching.

\section{Conclusions and outlook}

In this work, we have shown that appropriate choice of the dimensions of the cross resonator can achieve spectral overlap of the electric and magnetic resonances thus providing full 2$\pi$ phase coverage and high transmission amplitudes. We have employed eight types of cross resonators as building blocks to construct a spiraling phase plate with eight discretized phase levels. FDTD simulation results clearly show the eight discrete phase levels across the surface of the meta-surface optical element. There was a 12\% difference in the transmission amplitudes across the eight resonators observed. The metadevice was fabricated using standard projection micro-lithography and plasma etching. 

Such simple-to-fabricate optical elements are promising for the phase and polarisation control at the IR-THz beamline at the Australian Synchrotron. Applications in opto-mechanics also require high-transmission optical elements for imparting torque onto absorbing and scattering objects. The demonstrated simple fabrication method is applicable to a wider range of optical elements such as conical lenses (axicons), and gratings and can deliver higher quality surface finish as compared with 3D printing recently used for THz polarisation optics~\cite{17jopt}. Optical elements for spin and orbital angular momentum control are expected to provide new insights into  complex secondary structures of polymers including bio-polymers such as silk which have strong signatures at IR-THz spectral range~\cite{17m356,17sr7419,15a11863,17mre115028}.     
\\
\noindent
\section*{Acknowledgements}
This work was performed in part at the Melbourne Centre for Nanofabrication (MCN) in the Victorian Node of the Australian National Fabrication Facility (ANFF).

\bibliography{paper6b,report} 

\begin{thebibliography}{10}

\bibitem{yu2014flat}
Yu, N. and Capasso, F., ``Flat optics with designer metasurfaces,'' {\em Nature
  materials}~{\bf 13}(2),  139--150 (2014).

\bibitem{genevet2017recent}
Genevet, P., Capasso, F., Aieta, F., Khorasaninejad, M., and Devlin, R.,
  ``Recent advances in planar optics: from plasmonic to dielectric
  metasurfaces,'' {\em Optica}~{\bf 4}(1),  139--152 (2017).

\bibitem{sautter2015active}
Sautter, J., Staude, I., Decker, M., Rusak, E., Neshev, D.~N., Brener, I., and
  Kivshar, Y.~S., ``Active tuning of all-dielectric metasurfaces,'' {\em ACS
  nano}~{\bf 9}(4),  4308--4315 (2015).

\bibitem{chen2016review}
Chen, H.-T., Taylor, A.~J., and Yu, N., ``A review of metasurfaces: physics and
  applications,'' {\em Reports on Progress in Physics}~{\bf 79}(7),  076401
  (2016).

\bibitem{yu2015high}
Yu, Y.~F., Zhu, A.~Y., Paniagua-Dom{\'\i}nguez, R., Fu, Y.~H., Luk'yanchuk, B.,
  and Kuznetsov, A.~I., ``High-transmission dielectric metasurface with 2$\pi$
  phase control at visible wavelengths,'' {\em Laser \& Photonics Reviews}~{\bf
  9}(4),  412--418 (2015).

\bibitem{arbabi2015dielectric}
Arbabi, A., Horie, Y., Bagheri, M., and Faraon, A., ``Dielectric metasurfaces
  for complete control of phase and polarization with subwavelength spatial
  resolution and high transmission,'' {\em Nature nanotechnology}~{\bf 10}(11),
   937--943 (2015).

\bibitem{kildishev2013planar}
Kildishev, A.~V., Boltasseva, A., and Shalaev, V.~M., ``Planar photonics with
  metasurfaces,'' {\em Science}~{\bf 339}(6125),  1232009 (2013).

\bibitem{chong2015polarization}
Chong, K.~E., Staude, I., James, A., Dominguez, J., Liu, S., Campione, S.,
  Subramania, G.~S., Luk, T.~S., Decker, M., Neshev, D.~N., et~al.,
  ``Polarization-independent silicon metadevices for efficient optical
  wavefront control,'' {\em Nano letters}~{\bf 15}(8),  5369--5374 (2015).

\bibitem{khorasaninejad2016metalenses}
Khorasaninejad, M., Chen, W.~T., Devlin, R.~C., Oh, J., Zhu, A.~Y., and
  Capasso, F., ``Metalenses at visible wavelengths: Diffraction-limited
  focusing and subwavelength resolution imaging,'' {\em Science}~{\bf
  352}(6290),  1190--1194 (2016).

\bibitem{cheng2014wave}
Cheng, J., Ansari-Oghol-Beig, D., and Mosallaei, H., ``Wave manipulation with
  designer dielectric metasurfaces,'' {\em Optics letters}~{\bf 39}(21),
  6285--6288 (2014).

\bibitem{genevet2015holographic}
Genevet, P. and Capasso, F., ``Holographic optical metasurfaces: a review of
  current progress,'' {\em Reports on Progress in Physics}~{\bf 78}(2),  024401
  (2015).

\bibitem{yang2014dielectric}
Yang, Y., Wang, W., Moitra, P., Kravchenko, I.~I., Briggs, D.~P., and
  Valentine, J., ``Dielectric meta-reflectarray for broadband linear
  polarization conversion and optical vortex generation,'' {\em Nano
  letters}~{\bf 14}(3),  1394--1399 (2014).

\bibitem{khorasaninejad2015efficient}
Khorasaninejad, M., Zhu, W., and Crozier, K., ``Efficient polarization beam
  splitter pixels based on a dielectric metasurface,'' {\em Optica}~{\bf 2}(4),
   376--382 (2015).

\bibitem{chang2017demonstration}
Chang, C.-C., Headland, D., Abbott, D., Withayachumnankul, W., and Chen, H.-T.,
  ``Demonstration of a highly efficient terahertz flat lens employing tri-layer
  metasurfaces,'' {\em Optics Letters}~{\bf 42}(9),  1867--1870 (2017).

\bibitem{wang2015broadband}
Wang, Q., Zhang, X., Xu, Y., Tian, Z., Gu, J., Yue, W., Zhang, S., Han, J., and
  Zhang, W., ``A broadband metasurface-based terahertz flat-lens array,'' {\em
  Advanced Optical Materials}~{\bf 3}(6),  779--785 (2015).

\bibitem{niu2013experimental}
Niu, T., Withayachumnankul, W., Ung, B. S.-Y., Menekse, H., Bhaskaran, M.,
  Sriram, S., and Fumeaux, C., ``Experimental demonstration of reflectarray
  antennas at terahertz frequencies,'' {\em Optics express}~{\bf 21}(3),
  2875--2889 (2013).

\bibitem{cong2014highly}
Cong, L., Xu, N., Gu, J., Singh, R., Han, J., and Zhang, W., ``Highly flexible
  broadband terahertz metamaterial quarter-wave plate,'' {\em Laser \&
  Photonics Reviews}~{\bf 8}(4),  626--632 (2014).

\bibitem{wang2015ultrathin}
Wang, D., Gu, Y., Gong, Y., Qiu, C.-W., and Hong, M., ``An ultrathin terahertz
  quarter-wave plate using planar babinet-inverted metasurface,'' {\em Optics
  express}~{\bf 23}(9),  11114--11122 (2015).

\bibitem{ma2016terahertz}
Ma, Z., Hanham, S.~M., Albella, P., Ng, B., Lu, H.~T., Gong, Y., Maier, S.~A.,
  and Hong, M., ``Terahertz all-dielectric magnetic mirror metasurfaces,'' {\em
  ACS Photonics}~{\bf 3}(6),  1010--1018 (2016).

\bibitem{evlyukhin2010optical}
Evlyukhin, A.~B., Reinhardt, C., Seidel, A., Luk’yanchuk, B.~S., and
  Chichkov, B.~N., ``Optical response features of si-nanoparticle arrays,''
  {\em Physical Review B}~{\bf 82}(4),  045404 (2010).

\bibitem{17jopt}
Ryu, M., Linklater, D.~P., Hart, W., Bal\v{c}ytis, A., Skliutas, E.,
  Malinauskas, M., Appadoo, D., Tan, Y.-R.~E., Ivanova, E.~P., Morikawa, J.,
  and Juodkazis, S., ``{3D} printed polarising grids for {IR-THz} synchrotron
  radiation,'' {\em J. Optics} ,  (provisionally accepted) (2017).

\bibitem{17m356}
Bal\v{c}ytis, A., Ryu, M., Wang, X., Novelli, F., Seniutinas, G., Du, S., Wang,
  X., Li, J., Davis, J., Appadoo, D., Morikawa, J., and Juodkazis, S., ``Silk:
  Optical properties over 12.6 octaves {THz-IR-Visible-UV} range,'' {\em
  Materials}~{\bf 10}(4),  356 (2017).

\bibitem{17sr7419}
Ryu, M., Bal\v{c}ytis, A., Wang, X., Vongsvivut, J., Hikima, Y., Li, J., Tobin,
  M.~J., Juodkazis, S., and Morikawa, J., ``Orientational mapping augmented
  sub-wavelength hyper-spectral imaging of silk,'' {\em Sci. Reports}~{\bf 7},
  7419 (2017).

\bibitem{15a11863}
Morikawa, J., Ryu, M., Bal\v{c}ytis, A., Seniutinas, G., Fan, L., Mizeikis, V.,
  Li, J.~L., Wang, X.~W., Zamengo, M., Wang, X., and Juodkazis, S., ``Silk
  fibroin as water-soluble bio-resist and its thermal properties,'' {\em RSC
  Advances}~{\bf 6},  11863 -- 11869 (2015).

\bibitem{17mre115028}
Ryu, M., Kobayashi, H., Bal\v{c}ytis, A., Wang, X., Vongsvivut, J., Li, J.,
  Urayama, N., Mizeikis, V., Tobin, M., Juodkazis, S., and Morikawa, J.,
  ``Nanoscale chemical mapping of laser-solubilized silk,'' {\em Materials
  Research Express}~{\bf 4}(11),  115028 (2017).

\end{thebibliography}
\bibliographystyle{spiebib} 

\end{document}